\documentstyle[emulateapj,apjfonts,epsf]{article}

\received{}
\accepted{}
\journalid{}{}
\articleid{}{}

\slugcomment{Accepted by {\em The Astrophysical Journal Letters}
January 22, 2002}

\lefthead{Camilo et al.}
\righthead{PSR~J1124$-$5916 in SNR~G292.0+1.8}

\begin{document}

%
%

\def\psr{PSR~J1124$-$5916}
\def\snr{G292.0+1.8}
\def\chandra{{\em Chandra\/}}

\title{\psr: discovery of a young, energetic pulsar in the supernova
remnant \snr\ }

\author{F.~Camilo,\altaffilmark{1} 
  R.~N.~Manchester,\altaffilmark{2}
  B.~M.~Gaensler,\altaffilmark{3} 
  D.~R.~Lorimer,\altaffilmark{4} and
  J.~Sarkissian\altaffilmark{5} }
\medskip
\altaffiltext{1}{Columbia Astrophysics Laboratory, Columbia University,
  550 West 120th Street, New York, NY~10027}
\altaffiltext{2}{Australia Telescope National Facility, CSIRO,
  P.O.~Box~76, Epping, NSW~1710, Australia}
\altaffiltext{3}{Harvard-Smithsonian Center for Astrophysics, 60 Garden
  Street, Cambridge, MA~02138}
\altaffiltext{4}{University of Manchester, Jodrell Bank Observatory,
  Macclesfield, Cheshire, SK11~9DL, UK}
\altaffiltext{5}{Australia Telescope National Facility, CSIRO, Parkes
  Observatory, P.O.~Box~276, Parkes, NSW~2870, Australia}

\begin{abstract}
We report the discovery with the Parkes radio telescope of a pulsar
associated with the $\sim 1700$\,yr-old oxygen-rich composite supernova
remnant \snr.  \psr\ has period 135\,ms and period derivative
$7.4\times10^{-13}$, implying characteristic age 2900\,yr, spin-down
luminosity $1.2\times10^{37}$\,erg\,s$^{-1}$, and surface magnetic
field strength $1.0\times10^{13}$\,G.  Association between the pulsar
and the synchrotron nebula previously identified with \chandra\ within
this supernova remnant is confirmed by the subsequent detection of
X-ray pulsations by Hughes et~al.  The pulsar's flux density at
1400\,MHz is very small, $S \approx 80\,\mu$Jy, but the radio luminosity
of $S d^2 \sim 2$\,mJy\,kpc$^2$ is not especially so, although it is one
order of magnitude smaller than that of the least luminous young pulsar
previously known.  This discovery suggests that very deep radio
searches should be done for pulsations from pulsar wind nebulae in
which the central pulsed source is yet to be detected and possibly
other more exotic neutron stars.

\end{abstract}


\section{Introduction}\label{sec:intro}

The supernova remnant (SNR) \snr\ is one of only three oxygen-rich SNRs
known in the Galaxy.  The other two (Puppis~A and Cas~A) have central
compact objects, the nature of which however remains mysterious
(e.g.\ \cite{gbs00}; \cite{pza+00}). At radio wavelengths, \snr\ has
the appearance of a composite SNR, with a central peak and a shell
$\sim 10'$ in diameter (\cite{bgcr86}).  ASCA X-ray observations
(\cite{tts98}) detected a non-thermal nebula coincident with the
central radio component.  Recent \chandra\ (ACIS-S) observations have
shown this nebula to be $\sim 2'$ in extent, and to contain a resolved
compact source located near its peak.  This discovery, together with
the energetics of the nebula, provides nearly incontrovertible evidence
for the existence of a pulsar powering the nebula (\cite{hsb+01}).

The \chandra\ observations of \snr\ are a beautiful example of the
recent flood of X-ray data with high spatial, temporal and spectral
resolution that is advancing dramatically our understanding of the
varied outcomes of supernova explosions.  In particular, X-rays provide
an important complement to the radio band, the traditional hunting
ground of pulsar studies.

In this Letter we report the discovery and key parameters of the pulsar
in \snr\ in a deep observation with the Parkes radio telescope.  The
characterization of this young and energetic pulsar is important for
the analysis of existing X-ray and radio data on \snr.  Moreover, this
discovery has significant implications for the concept of ``radio-quiet
neutron stars.''

\section{Observations}\label{sec:obs}

The location of \snr\ was searched by the Parkes Multibeam Pulsar
Survey of the inner Galactic plane ($260\arcdeg<l<50\arcdeg$;
$|b|<5\arcdeg$) that has discovered more than 600 pulsars \linebreak

\medskip
\epsfxsize=8.0truecm
\epsfbox{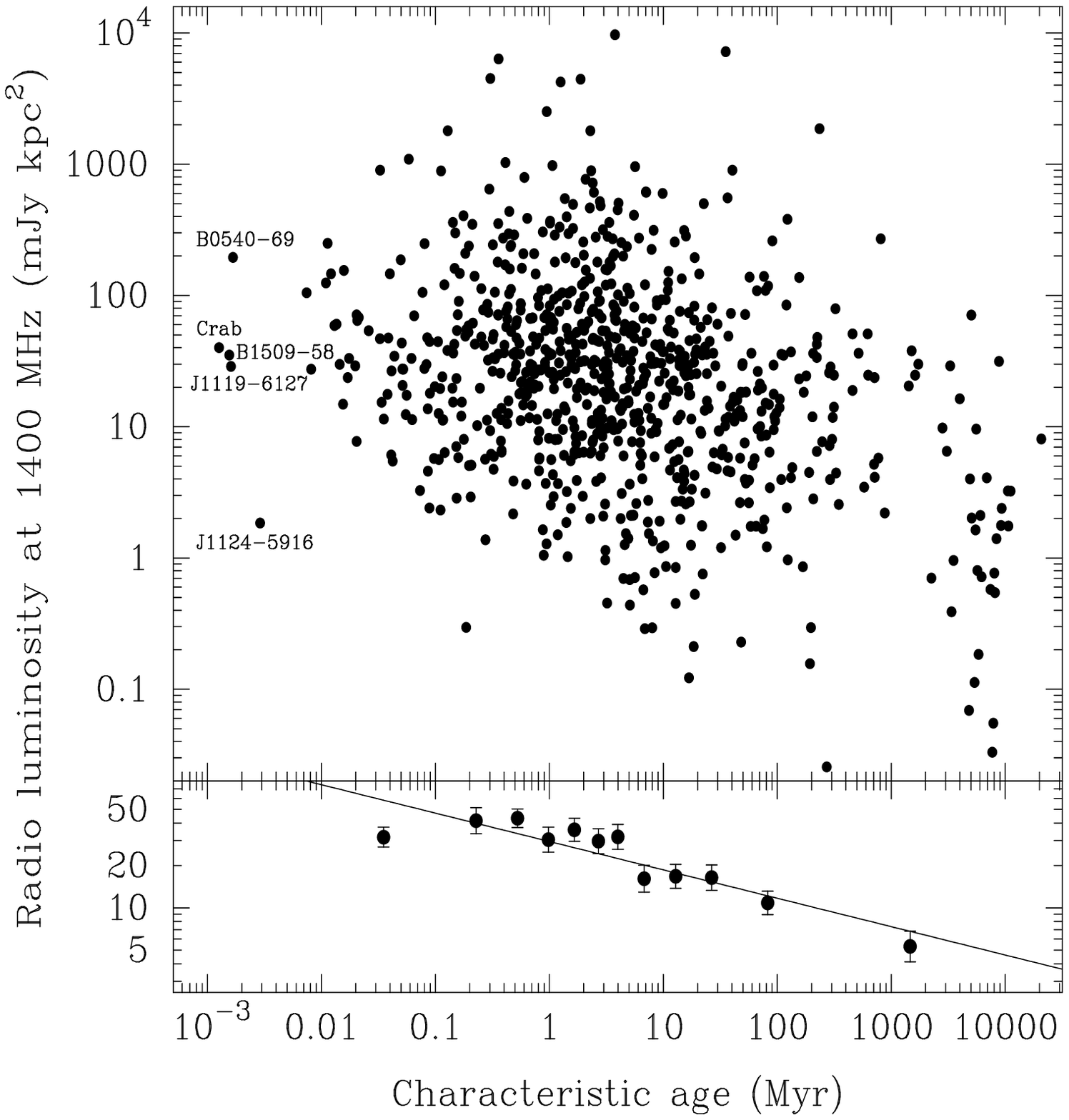}
\figcaption[f1.eps]{\label{fig:Lr}
{\em Top\/}: Radio luminosities at 1400\,MHz ($L_{1400}$) for 828
pulsars plotted vs. characteristic age $\tau_c$.  About 5\% of known
pulsars have $L_{1400}$ below that of \psr.  {\em Bottom\/}: Data from
{\em top panel\/} binned into 12 groups of 69 pulsars each, with points
representing the mean of $L_{1400}$ and $\tau_c$ in each bin, and a
straight line least-squares fit to the binned points.  The left-most
point ($\tau_c \la 100$\,kyr) suggests that young pulsars are not
particularly luminous by comparison with middle-aged pulsars
($100\,\mbox{kyr} \la \tau_c \la 5$\,Myr).  Some luminosity decay for
larger ages is suggested by the straight-line fit.  }

\bigskip

\noindent (\cite{mlc+01}).  The flux density limit at the location of the
\chandra\ pulsar candidate was $S_\nu \approx 0.3$\,mJy at a frequency
$\nu = 1374$\,MHz.  A similar limit was reached in a directed search of
the SNR by Kaspi et al.~(1996)\nocite{kmj+96} at $\nu = 1520$\,MHz.  At
a distance $d \sim 5$\,kpc (see below), the corresponding luminosity
limit is a not particularly constraining $L_{1400} \equiv S_{1400} d^2
\sim 7$\,mJy\,kpc$^2$: $\sim 25\%$ of known pulsars have luminosities
below this value (see Fig.~\ref{fig:Lr}).


On September 5, 2001 we searched the pulsar candidate position (see
Table~\ref{tab:parms}) at the Parkes telescope using the center beam of
the multibeam receiver at a central frequency of 1374\,MHz with a
filter bank spectrometer having 96 channels spanning a total band of
288\,MHz for each of two polarizations.  Signals from each channel and
polarization were square-law detected, high-pass filtered and summed in
polarization pairs. Each of the 96 channels was then integrated and
one-bit digitized at 1\,ms intervals and recorded to magnetic tape for
off-line analysis. The total observation time was 10.2\,hr.


Standard search algorithms were used to reduce the data.  First, strong
narrow-band radio interference was identified and removed by masking
five offending frequency channels.  Next, data from $2^{25}$ time
samples (9.3\,hr) were de-dispersed at 325 trial dispersion measures
($0 \le \mbox{DM} \le 8800$\,cm$^{-3}$\,pc).  Each of the 325 resulting
time series was then searched for periodic signals over a range of duty
cycles with an FFT-based code (described in detail by \cite{lkm+00})
identifying significant features in the fundamental amplitude spectrum
as well as in spectra with 2, 4, 8, and 16 harmonics folded in.
Finally a limited but finer search in period and DM was made of the
entire data set for the best dispersed candidates from this stage.  A
good pulsar candidate was found with signal-to-noise ratio
$\mbox{S/N}=12.4$, barycenter-corrected period $P=135.3126$\,ms, and
$\mbox{DM}=330$\,cm$^{-3}$\,pc.

On October 3, 2001 we confirmed the pulsar with a detection of
$\mbox{S/N}=8.8$ in a 5\,hr observation, with identical DM and
$P=135.3144$\,ms, immediately implying a large period derivative $\dot
P = 7.4 \times 10^{-13}$.  Following the initial radio detection, new
\chandra\ (HRC) data were searched for X-ray pulsations.  As detailed
by Hughes et al. (2002) in a companion Letter, pulsations were found at
a consistent period from the location of the previously identified
pulsar wind nebula (PWN), ensuring beyond any doubt that we have
discovered the pulsar near the center of SNR~\snr.

The measured DM, together with the Taylor \& Cordes (1993)\nocite{tc93}
model for the Galactic distribution of free electrons, implies a pulsar
distance of 11\,kpc.  However the pulsar is located in the direction of
the Carina spiral arm, where the model suffers from systematic errors.
In fact, a new electron density/distance model (Cordes \& Lazio, in
preparation) suggests a pulsar distance of 5.7\,kpc (J.~Cordes, private
communication).  The SNR shows H{\sc i} absorption to beyond the
tangent point in this direction (\cite{cmr+75}), implying a lower limit
on the distance of 3.2\,kpc.  Since the SNR shows absorption towards
all more distant emission, no upper limit can be put on its distance
from these data.  A distance estimate is also available from the
observed value of reddening, $d \sim 5.4$\,kpc (\cite{gsz+79}).  In
light of these various constraints on pulsar/SNR distance we adopt
hereafter the estimate $d = 5$\,kpc.

We have begun regular timing measurements of the new pulsar at Parkes,
with 3--5\,hr typically needed to obtain a satisfactory pulse profile
and corresponding time-of-arrival (TOA).  The average pulse profile has
a single approximately symmetric component with $\mbox{FWHM} \sim 0.1
P$.  Using the TOAs, the arcsec-accuracy position measured with
\chandra, and the {\sc tempo}\footnote{See
http://pulsar.princeton.edu/tempo.} timing software, we have obtained a
phase-connected solution, with the resulting spin parameters listed in
Table~\ref{tab:parms}. The pulsar appears to suffer a large amount of
``timing noise,'' with large and systematic residuals from the
second-order fit listed in Table~\ref{tab:parms}. Fitting of four
spin-frequency derivatives removed the systematic component of this
noise, leaving a final rms timing residual of 0.4\,ms. The S/Ns have
remained constant when scaled by integration time, suggesting that the
pulsar's flux is not greatly modulated by interstellar scintillation.
We use this and the known observing parameters, telescope gain and
system temperature, including the contribution to the latter from the
SNR (\cite{lgcm77}), to estimate a flux density of $80 \pm 20\,\mu$Jy.

\section{Discussion}\label{sec:disc}

The measured $P$ and $\dot P$ of \psr\ imply that it is a very young
and energetic pulsar.  The implied characteristic age and surface
magnetic dipole field strength are $\tau_c = P/2\dot P = 2900$\,yr and
$B = 3.2\times10^{19} (P \dot P)^{1/2} = 1.0 \times 10^{13}$\,G
respectively.  The spin-down energy loss rate $\dot E = 4\pi^2 I \dot P
/P^3 = 1.2 \times10^{37}$\,erg\,s$^{-1}$ (where a neutron star moment
of inertia $I = 10^{45}$\,g\,cm$^2$ has been used), in excellent
agreement with the value predicted by Hughes et al.
(2001)\nocite{hsb+01} from the energetics of the PWN, if $d \sim
5$\,kpc.  In comparison with the sample of $\sim 1500$ rotation-powered
pulsars known, J1124$-$5916 ranks as the 6th youngest in terms of
$\tau_c$ and the 8th most energetic in terms of $\dot E$.

The age of \snr\ is derived to be $\la 1700\,(d/5\,\mbox{kpc})$\,yr
from a measurement of the high radial velocity oxygen-rich material
positionally coincident with the central synchrotron nebula
(\cite{mc79}; \cite{bgdb83}), while the pulsar characteristic age is a
factor of nearly two larger than this.  Assuming that the age estimate
for the SNR represents the true age of the system, this discrepancy can
be simply explained if the initial spin period of \psr\ was $P_0 \ga
90$\,ms (e.g.\ \cite{krv+01}; \cite{mss+01}).  This is slower than for
the six other young pulsars whose initial periods have been estimated,
all of which have $P_0 < 60$\,ms.  Alternatively the pulsar may spin
down with a braking torque that is different from the usually assumed
constant magnetic dipole.  In a study of pulsar population dynamics,
Cordes \& Chernoff (1998)\nocite{cc98} suggest alternate forms of
braking torque (e.g.\ a braking index of 2.5 with a magnetic field
decay time $\sim 6$\,Myr) that result in actual pulsar ages smaller
than $\tau_c$ by a factor $\la 2$.  In the absence of a measured
braking index for \psr, we tentatively regard a relatively large $P_0$
as a preferred explanation.

In Table~\ref{tab:young} we summarize key parameters for the youngest
rotation-powered pulsars known, ordered by decreasing $\dot E$.
Heading the list are the three ``Crab-like'' pulsars, with extremely
large $\dot E$.  The remaining six young pulsars are varied in their
properties and generally follow the trend of increasing $P$ and $B$
with decreasing $\dot E$ (with the pulsar in G11.2$-$0.3 a notable
exception).  Their spin-down luminosities are 20--200 times lower than
the Crab's and many are therefore difficult to detect: four were
discovered just in the past two years (two each in radio and X-rays).


In its spin parameters \psr\ is most similar to PSR~B1509$-$58 (see
Table~\ref{tab:young}).  It is interesting to note that their
respective PWNe also have comparable luminosities.  In the 0.2--4\,keV
X-ray band, we find for the PWN powered by J1124$-$5916
that\footnote{The photon index of this source could not be measured
from the \chandra\ observations of Hughes et al. (2001\nocite{hsb+01})
because of contamination by thermal emission from the SNR.  We assume a
photon index $\Gamma = 2$, as is typical for such sources.} $L_X =
6\times10^{34}$\,erg\,s$^{-1} = 0.005 \dot{E}$ for a distance 5\,kpc
(\cite{hsb+01}), while for B1509$-$58 we find $L_X =
2\times10^{35}$\,erg\,s$^{-1} = 0.01 \dot{E}$ in the same energy range
(\cite{gak+01}).  For their respective radio PWNe, we find $L_R =
4\times10^{33}$\,erg\,s$^{-1} = 0.0003 \dot{E}$ for J1124$-$5916
(\cite{hsb+01}), while $L_R \sim 5\times10^{33}$\,erg\,s$^{-1} = 0.0003
\dot{E}$ for B1509$-$58 (\cite{gak+01}).

Thus the efficiencies with which these pulsars power their X-ray
nebulae differ only by a factor of $\sim2$, while those for their radio
nebulae are approximately equal.  Although these pulsars have
comparable spin parameters, we explain below that their PWNe have quite
different environments and likely evolutionary histories. It is
therefore surprising that the efficiencies with which they convert
their spin-down into nebular flux are so similar.

In the radio band, it has been argued that a pulsar's initial
parameters can make a crucial difference to the resultant luminosity of
its PWN at later times (e.g.\ \cite{bha90}).  If a pulsar is born
spinning rapidly and has a high magnetic field, it will dump most of
its spin-down energy into its surrounding nebula at the earliest
stages, when adiabatic losses are most severe; this will result in a
comparatively underluminous radio PWN (\cite{bha90}; \cite{cgk+01}).
However, a pulsar that is otherwise similar but born spinning more
slowly will not release its energy so quickly, so that its nebula will
suffer less from expansion losses and will be correspondingly brighter
at radio wavelengths. We have argued above that \psr\ was possibly born
with a comparatively long spin period, $P_0 \ga 90$\,ms, while
PSR~B1509$-$58 is generally assumed to have been born with a Crab-like
$P_0 \approx 20$\,ms (\cite{bha90}).  The comparable values of $L_R$
for their PWNe are therefore not easily explained.

Furthermore, the fact that the radio and X-ray extents of the PWN
powered by PSR~B1509$-$58 are approximately equal indicates that
synchrotron cooling does not yet dominate the nebula at high energies
(\cite{gak+01}), and can account for this nebula's comparatively low
X-ray efficiency (\cite{che00}). However, for \psr\ the X-ray nebula is
noticeably smaller than its radio counterpart (\cite{hsb+01}; Gaensler
\& Wallace, in preparation), indicating that the X-ray emitting
electrons are in this case efficient radiators.  Thus we expect the
X-ray luminosity of the PWN around J1124$-$5916 to be much larger than
observed, closer to the factor $L_X \approx 0.05 \dot{E}$ seen for the
Crab Nebula and other X-ray PWNe dominated by radiative losses
(\cite{che00}).

Thus both $L_R$ and $L_X$ for the PWN powered by \psr\ are much lower
than expected through simple physical arguments.  There are a variety
of other factors which can affect a PWN's luminosity: the pulsar
braking index, PWN magnetic field, and nebular expansion velocity all
are important parameters at radio wavelengths (\cite{rc84};
\cite{cgk+01}), while the Lorentz factor of the wind, the radius of the
termination shock and the magnetization parameter, $\sigma$, all have a
strong bearing on the nebula's X-ray luminosity (\cite{che00}).  The
unexpectedly similar nebular efficiencies for PSRs~J1124$-$5916 and
B1509$-$58, along with the large range of efficiencies among the other
young pulsars in Table~\ref{tab:young}, emphasize that we still lack a
detailed understanding of how a pulsar ultimately deposits its energy.

One particularly interesting young pulsar in this regard is
PSR~J1119$-$6127 (\cite{ckl+00}; Table~\ref{tab:young}), which has no
known radio-bright PWN down to extremely constraining surface
brightness limits (\cite{cgk+01}).  It also has no confirmed
X-ray-bright PWN (although a possible candidate was detected by
\cite{pkc+01}).  This example, together with several relatively young
and energetic pulsars that do not power detectable PWNe
(\cite{gsf+00}), suggests that pulsar environments need not be good
``calorimeters.''  Thus, some young pulsars may pass unnoticed unless
their beamed radiation is favorably oriented so that pulsations may be
detected.

That radio pulsar beams do not in general sweep $4\pi$\,sr is a fact of
life.  Although it is generally accepted that the beaming fraction $f$
is period dependent, with short-period pulsars beaming to a larger
fraction of the sky than their long-period counterparts, a consensus
has yet to emerge on the form of $f(P)$.  Pulse width analyses suggest
$f \approx 0.3$ for a pulsar with $P \sim 0.1$\,s (\cite{tm98}) while,
from an analysis of pulsar--PWN associations, Frail \& Moffett
(1993)\nocite{fm93} find $f \approx 0.6$. Since PWNe are regarded as
unambiguous indicators that a young pulsar is present, the failure to
detect pulsations towards such sources is usually ascribed solely to a
misoriented pulsar beam.  However, the case of \snr\ and its faint
radio pulsar highlights an important caveat: the radio
luminosity\footnote{The radio luminosities we discuss here are really
``pseudo-luminosities'': they assume that total luminosity is
proportional to the integrated flux density of the observed cut across
the radio beam.  A realistic discussion of actual luminosities depends
crucially on the generally unknown pulsar beam shape: e.g., for a beam
with exponential roll off, the measured flux density --- and therefore
pseudo-luminosity --- can be arbitrarily low depending on the viewing
impact angle, even for a large beam-averaged luminosity.  Here
primarily we are concerned with empirically determined
``luminosities,'' i.e., with pseudo-luminosities.} limit implied by a
non-detection should now be below at least $L_{1400} \sim
1$\,mJy\,kpc$^2$, approximately the luminosity of \psr, and possibly as
low as $\la 0.1$\,mJy\,kpc$^2$ (see Fig.~\ref{fig:Lr}), before one can
invoke unfavorable beaming as an explanation.

Various analyses (e.g.\ \cite{ec89}) have suggested that young pulsars
have higher radio luminosities than older ones.  Figure~\ref{fig:Lr}
shows that the basis for such arguments is weak in the case of very
young pulsars.  The least luminous radio pulsar previously detected for
which $\tau_c < 10$\,kyr had $L_{1400} \sim 30$\,mJy\,kpc$^2$.  The
fact that some X-ray-detected pulsars have had much more constraining
luminosity limits put on their radio emission (e.g.\ \cite{ckm+98};
Table~\ref{tab:young}) has usually been taken to imply that these
pulsars have radio beams which are not directed towards us.  However,
the discovery of \psr\ shows that very young pulsars can certainly have
much lower radio luminosities than previously thought, either due to
intrinsically low luminosity or to our line of sight cutting only
low-level wings on the emitted beam.

The above discussion serves as a cautionary tale in the context of
discussions of exotic sources such as the ``radio-quiet neutron stars''
(RQNSs), ``anomalous X-ray pulsars'' (AXPs) and ``soft $\gamma$-ray
repeaters'' (SGRs). While there is no question that these sources have
properties which make them distinct from ``normal'' radio pulsars
(e.g.\ \cite{gpv99}; \cite{pkc00}; Gaensler et al.~2000a\nocite{gbs00};
\cite{kgc+01}; \cite{cph+01}), the fact that none of these sources has
been detected at radio wavelengths has been taken to imply that their
radio pulse mechanism must be inactive (\cite{bh98b}; \cite{zh00};
\cite{zha01}).  However, the radio luminosity limits for many of these
sources are not particularly stringent by the standard of radio
pulsars, and there are so few of these objects known that they all
could be radio luminous and yet all could still be beaming away from us
(\cite{gsgv01}).  It may well be that the RQNSs, AXPs and SGRs are all
truly radio-silent, but this conclusion is difficult to justify until
the luminosity limits available for such objects reach a level
equivalent to at least the smallest radio luminosities observed for
radio pulsars, $L_{1400} \sim 0.1$\,mJy\,kpc$^2$ (Fig.~\ref{fig:Lr}),
and in principle even smaller values.

With the discovery of J1124$-$5916, \snr\ becomes the second example of
an oxygen-rich SNR with a confirmed rapidly spinning pulsar (after
0540$-$693 in the LMC).  It remains to be seen what is the true nature
of the compact sources in the other two Galactic oxygen-rich SNRs.
This discovery also suggests that very deep searches for radio
pulsations from known PWNe and possibly other more exotic neutron stars
may be well worthwhile.

\acknowledgments

We are deeply grateful to John Reynolds for his support of this
project.  The Parkes radio telescope is part of the Australia Telescope
which is funded by the Commonwealth of Australia for operation as a
National Facility managed by CSIRO.  FC acknowledges support from NASA
grant NAG5-9095.  BMG is supported by a Clay Fellowship awarded by the
Harvard-Smithsonian Center for Astrophysics.  DRL is a University
Research Fellow funded by the Royal Society.

\clearpage

%
%
\begin{deluxetable}{ll}
\tablecaption{\label{tab:parms}Parameters of \psr\ }
\tablecolumns{2}
\tablewidth{0pc}
\tablehead{
\colhead{Parameter}    &
\colhead{Value}     \nl}
\startdata
R.A. (J2000)\tablenotemark{a}\dotfill          & $11^{\rm h}24^{\rm m}39\fs1$\nl
Decl. (J2000)\tablenotemark{a}\dotfill         & $-59\arcdeg16'20''$         \nl
Period, $P$ (s)\dotfill                        & 0.1353140749(2)             \nl
Period derivative, $\dot P$\dotfill            & $7.471(2)\times 10^{-13}$   \nl
Epoch (MJD)\dotfill                            & 52180.0                     \nl
Dispersion measure, DM (cm$^{-3}$\,pc)\dotfill & 330(2)                      \nl
Data span (MJD)\dotfill                        & 52157--52214                \nl
Number of TOAs\dotfill                         & 10                          \nl
Rms timing residual (ms)\dotfill               & 5.8                         \nl
Rms timing residual, ``whitened'' (ms)\dotfill & 0.4                         \nl
Flux density at 1400\,MHz, $S_{1400}$ (mJy)\dotfill        
                                               & 0.08(2)                     \nl
Derived parameters:                                      &                   \nl
~~Characteristic age, $\tau_c$ (yr)\dotfill              & 2900              \nl
~~Spin-down luminosity, $\dot E$ (erg\,s$^{-1}$)\dotfill & $1.2\times10^{37}$\nl
~~Magnetic field strength, $B$ (G)\dotfill               & $1.0\times10^{13}$\nl
~~Distance, $d$ (kpc)\tablenotemark{b}\dotfill           & $\sim 5$          \nl
~~Radio luminosity, $L_{1400}$ (mJy\,kpc$^2$)\dotfill    & $\sim 2$          \nl
\enddata
\tablecomments{Numbers in parentheses represent 1\,$\sigma$
uncertainties in the least-significant digits quoted.}
\tablenotetext{a}{Position known with $\sim 1''$ accuracy from
\chandra\ data (\cite{hsb+01}).}
\tablenotetext{b}{Distance of SNR~\snr\ (see text).}
\end{deluxetable}

\begin{deluxetable}{lccccllr}
\tablecaption{\label{tab:young}Pulsars with apparent age $< 5$\,kyr }
\tablecolumns{8}
\tablewidth{0pc}
\tablehead{
\colhead{PSR}          &
\colhead{$\tau_c$}     &
\colhead{$P$}          &
\colhead{$B$}          &
\colhead{$\dot E$}     &
\colhead{SNR}          &
\colhead{$d$}          &
\colhead{$L_{1400}$} \nl
\colhead{}                           &
\colhead{(kyr)}                      &
\colhead{(ms)}                       &
\colhead{($10^{12}$\,G)}             &
\colhead{($10^{36}$\,erg\,s$^{-1}$)} &
\colhead{}                           &
\colhead{(kpc)}                      &
\colhead{(mJy\,kpc$^2$)}          \nl} 
\startdata
J0537$-$6910 &  5.0 &  16 &  1 & 480 & N157B        & 50 (LMC) &   $< 150 $  \nl
B0531$+$21   &  1.3 &  33 &  4 & 440 & Crab         &  2.0     &       40    \nl
B0540$-$69   &  1.7 &  50 &  5 & 150 & 0540$-$693   & 50 (LMC) &      200    \nl
J0205$+$6449 &  5.4 &  65 &  4 &  27 & 3C~58        &  2.6     &   $\la 2 $  \nl
B1509$-$58   &  1.6 & 150 & 15 &  18 & G320.4$-$1.2 &  5.2     &       35    \nl
J1124$-$5916 &  2.9 & 135 & 10 &  12 & G292.0+1.8   & $\sim 5$ &  $\sim 2 $  \nl
J1846$-$0258 &  0.7 & 323 & 48 &   8 & Kes~75       & $\sim19$ &  $\la 50 $  \nl
J1811$-$1926 & 24   &  65 &  2 &   6 & G11.2$-$0.3  &  5       &     $< 2 $  \nl
J1119$-$6127 &  1.6 & 407 & 41 &   2 & G292.2$-$0.5 & $\sim 6$ & $\sim 30 $  \nl
\enddata
\tablecomments{We list in order of decreasing spin-down luminosity
$\dot E$ all known pulsars for which the characteristic age $\tau_c
<5$\,kyr or for which the pulsar is possibly associated with a
historical supernova (J0205+6449: SN1181; J1811$-$1926: SN386). }
\end{deluxetable}

\end{document}